\documentclass[12pt]{article}
\usepackage{graphicx}
\usepackage{lscape}
\usepackage{citesort}

\begin{document}

\def\ds{\displaystyle}
\def\beq{\begin{equation}}
\def\eeq{\end{equation}}
\def\bea{\begin{eqnarray}}
\def\eea{\end{eqnarray}}
\def\beeq{\begin{eqnarray}}
\def\eeeq{\end{eqnarray}}
\def\ve{\vert}
\def\vel{\left|}
\def\ver{\right|}
\def\nnb{\nonumber}
\def\ga{\left(}
\def\dr{\right)}
\def\aga{\left\{}
\def\adr{\right\}}
\def\lla{\left<}
\def\rra{\right>}
\def\rar{\rightarrow}
\def\nnb{\nonumber}
\def\la{\langle}
\def\ra{\rangle}
\def\ba{\begin{array}}
\def\ea{\end{array}}
\def\tr{\mbox{Tr}}
\def\ssp{{\Sigma^{*+}}}
\def\sso{{\Sigma^{*0}}}
\def\ssm{{\Sigma^{*-}}}
\def\xis0{{\Xi^{*0}}}
\def\xism{{\Xi^{*-}}}
\def\qs{\la \bar s s \ra}
\def\qu{\la \bar u u \ra}
\def\qd{\la \bar d d \ra}
\def\qq{\la \bar q q \ra}
\def\gGgG{\la g^2 G^2 \ra}
\def\q{\gamma_5 \not\!q}
\def\x{\gamma_5 \not\!x}
\def\g5{\gamma_5}
\def\sb{S_Q^{cf}}
\def\sd{S_d^{be}}
\def\su{S_u^{ad}}
\def\ss{S_s^{??}}
\def\sbp{{S}_Q^{'cf}}
\def\sdp{{S}_d^{'be}}
\def\sup{{S}_u^{'ad}}
\def\ssp{{S}_s^{'??}}
\def\sig{\sigma_{\mu \nu} \gamma_5 p^\mu q^\nu}
\def\fo{f_0(\frac{s_0}{M^2})}
\def\ffi{f_1(\frac{s_0}{M^2})}
\def\fii{f_2(\frac{s_0}{M^2})}
\def\O{{\cal O}}
\def\sl{{\Sigma^0 \Lambda}}
\def\es{\!\!\! &=& \!\!\!}
\def\ap{\!\!\! &\approx& \!\!\!}
\def\ar{&+& \!\!\!}
\def\ek{&-& \!\!\!}
\def\kek{\!\!\!&-& \!\!\!}
\def\cp{&\times& \!\!\!}
\def\se{\!\!\! &\simeq& \!\!\!}
\def\eqv{&\equiv& \!\!\!}
\def\kpm{&\pm& \!\!\!}
\def\kmp{&\mp& \!\!\!}


\def\simlt{\stackrel{<}{{}_\sim}}
\def\simgt{\stackrel{>}{{}_\sim}}


\title{
         {\Large
                 {\bf
$B \rar K_2 \ell^+ \ell^-$ decay beyond the Standard Model
                 }
         }
      }

\author{\vspace{1cm}\\
{\small T. M. Aliev \thanks {e-mail:
taliev@metu.edu.tr}~\footnote{permanent address:Institute of
Physics,Baku,Azerbaijan}\,\,, M. Savc{\i} \thanks
{e-mail: savci@metu.edu.tr}} \\
{\small Physics Department, Middle East Technical University,
06531 Ankara, Turkey }\\
       }

\date{}

\begin{titlepage}
\maketitle
\thispagestyle{empty}

\begin{abstract}
The exclusive $B \rar K_2 \ell^+ \ell^-$ decay is studied using the most
general, model independent four--fermion interaction. The sensitivity of
the ratio of the decay widths when $K_2$ meson is longitudinally and
transversally polarized, the forward--backward asymmetry and longitudinal
polarization of the final lepton on the new Wilson coefficients is studied.
It is found that these quantities are very useful for establishing new
physics beyond the Standard Model. 
\end{abstract}

~~~PACS numbers: 12.60.--i, 13.20.--v, 13.20.He
\end{titlepage}

\section{Introduction}
Flavor changing neutral current (FCNC) processes provide promising
direction for testing the gauge structure of the Standard Model (SM)
at loop level, as these decays are forbidden at tree level. Moreover, FCNC
decays are sensitive to the new physics beyond the SM. Among all the FCNC
processes the rare $B$--meson decays occupy an important place since they
contain rich phenomena relevant to the SM and new physics beyond it. The
rare decays due to $b \to s$ transition receives special attention since SM
predicts ``large" branching ratio for them.

The radiative $B \to K^* \gamma$ \cite{R10701,R10702,R10703}, $B \to K_1
(1270,1430) \gamma$ \cite{R10704} and semileptonic $B \to K^* (892) \ell^+
\ell^-$ \cite{R10705,R10706} have been measured in experiments. 
\cite{R10704}. The isospin and forward--backward asymmetry in 
$B \to K^* (892) \ell^+ \ell^-$ is also measured by BaBar Collaboration 
\cite{R10707,R10708}. The semileptonic decays $B \to K (K^*) \ell^+ \ell^-$
theoretically are studied intensively in literature \cite{R10711}. The
radiative $B \to K_2 (1430) \gamma$ decay has also been observed at BaBar
\cite{R10709,R10710}.

The exclusive $B \to K_2 (1430) \ell^+ \ell^-$ decay is studied within the
SM in \cite{R10712,R10713} and it is obtained that the branching ratio of
this channel is comparable with that of the $B \to K^* (892) \ell^+ \ell^-$ 
decay. Therefore, investigation of the $B \to K_2 (1430) \ell^+ \ell^-$
decay can provide an independent test of the SM.
The main similarity of the $B \to K^* (892) \ell^+ \ell^-$ and
$B \to K_2 (1430) \ell^+ \ell^-$ decays in the SM is that both decays are
described by the $b\to s$ transition, hence by the same effective
Hamiltonian. 

In SM the $b \to s \ell^+ \ell^-$ transition is described by
the Wilson coefficients $C_7$, $C_9$ and $C_{10}$ of the operators
${\cal O}_7$, ${\cal O}_9$ and ${\cal O}_{10}$ at $\mu=m_b$ scale. Explicit 
expressions of these Wilson
coefficients and relevant operators can be found in \cite{R10714}. The new
physics in FCNC transitions can appear in two different ways: a) By new
contributions to the Wilson coefficients, i.e., by modification of the
Wilson coefficients, b) appearance of new operators which are absent in the
SM.

It should be noted that, the most general model independent analysis of the
$b \to s \ell^+ \ell^-$ transition is carried out in \cite{R10715} in terms
of ten new types of local four--Fermi interaction. Furthermore, extensive
study on various observables of the $b \to s \ell^+ \ell^-$ transition is
performed in the same framework in \cite{R10716,R10717}.
Extension to the
exclusive $B \to K(K^*) \ell^+ \ell^-$ channels are performed in
\cite{R10718,R10719,R10720}.

One important quantity in checking the predictions of the SM and
establishing new physics is the lepton polarization effects, which are first
pointed out in \cite{R10721} and subsequently considered in many works (see
for example references in \cite{R10722}). The main goal of the present work
is to study the branching ratio, forward--backward asymmetry and the final
lepton polarization effects for the rare $B \to K_2 (1430) \ell^+
\ell^-$decay in a model independent manner. It should be noted here that, $B
\to K_2(\to K \pi) \ell^+ \ell^-$ decay is studied in the SM and in the two 
new physics scenarios, namely vector--like quark model and the familiar 
non--universal $Z^\prime$ model in \cite{R10723}. 
The work is arranged as follows. In section 2, we firstly present the most 
general form of the local four--Fermi interactions and then using this form
we calculate the helicity amplitudes for the $B \to K_2 (1430) \ell^+ \ell^-$
decay. In this section we also present the expressions of the branching
ratio, forward--backward asymmetry and lepton polarizations in terms of
helicity amplitudes. Section 3 contains our numerical analysis and
conclusions.

\section{Helicity amplitudes for the $B \to K_2 \ell^+ \ell^-$ decay}

The $B \to K_2 \ell^+ \ell^-$ decay is described at quark level by the 
$b \to s \ell^+ \ell^-$ transition. After integrating over heavy degrees of
freedom, the effective Hamiltonian in the SM for the $b \to s \ell^+ \ell^-$
transition can be expressed as,
\bea
\label{e10701}
{\cal H}_{eff} = - {4 G_F \over \sqrt{2}} V_{tb} V_{ts}^* \sum_{i=1}^{10}
C_i(\mu) {\cal O}_i(\mu)~,
\eea
where ${\cal O}_i(\mu)$ are the four--quark operators and $C_i(\mu)$ are the
corresponding Wilson coefficients at $\mu$ scale. Explicit expressions of
the Wilson coefficients at next--to--next leading logarithm (NNLL) are
calculated in many works (for example see \cite{R10724} and the references
therein). The operators responsible for for the $B \to K_2 \ell^+ \ell^-$
decay are ${\cal O}_7$, ${\cal O}_9$ and ${\cal O}_{10}$ are given as,
\bea
\label{e10702}
{\cal O}_7 \es {e^2 \over 16 \pi^2} m_b (\bar{s}_R \sigma_{\mu\nu} b_R)
\, F^{\mu\nu}~, \nnb \\
{\cal O}_9 \es {e^2 \over 16 \pi^2} (\bar{s}_L \gamma_\mu b_L)
\, \bar{\ell} \gamma_\mu \ell~, \nnb \\
{\cal O}_{10} \es {e^2 \over 16 \pi^2} (\bar{s}_L \gamma_\mu b_L)     
\, \bar{\ell} \gamma_\mu \gamma_5 \ell~,
\eea
where $q_{L(R)}= \ds {1 \mp \gamma_5 \over 2} q$. Using this effective
Hamiltonian, the matrix element for the $b \to s \ell^+ \ell^-$ transition
can be written as,
\bea
\label{e10703}
{\cal M} \es {G _F \over \sqrt{2}\pi} V_{tb} V_{ts}^* \Bigg\{
C_9^{eff} \bar{s}_L \gamma_\mu b_L \, \bar{\ell} \gamma_\mu \ell +
C_{10} \bar{s}_L \gamma_\mu b_L \, \bar{\ell} \gamma_\mu \gamma_5 \ell
- {2C_7^{eff} \over q^2} \bar{s}_R i \sigma_{\mu\nu} q^\nu b_R \, \bar{\ell}
\gamma_\mu \ell \Bigg\}~,
\eea
where $C_9^{eff}$ contains short and long distance contributions and it can
be written as,
\bea
\label{e10704}
C_9^{eff} = C_9(\mu) + Y_{SD}(z,\hat{s}) + Y_{LD}(z,\hat{s})~.
\eea
Here, $z=m_c/m_b$, $\hat{s}=q^2/m_b^2$, $Y_{SD}$ describes contributions
coming from four--quark operators and $Y_{LD}$ corresponds to the long
distance effects from four--quark operators near the $\bar{c}c$ resonance.
The expressions for $Y_{SD}$ and $Y_{SD}$ can be written as
\bea 
\label{e10705}
Y_{SD}(z,\hat{s}) \es h \ga z,\hat s \dr
C^{(0)}(\mu)
- \frac{1}{2} h \ga 1,\hat s \dr
\left[4 C_3(\mu) + 4 C_4(\mu) + 3 C_5(\mu) + C_6(\mu) \right] \nnb \\
\ek \frac{1}{2} h \ga 0,\hat s \dr
\left[ C_3(\mu) + 3  C_4(\mu) \right]
+ \frac{2}{9} \left[ 3 C_3(\mu) + C_4(\mu) + 3 C_5(\mu) + C_6(\mu) \right]~,
\eea
where $x=4 z^2/\hat s$, and, 
\bea
\label{e10706}
C^{(0)}(\mu) \es 3 C_1(\mu) + C_2(\mu) + 3 C_3(\mu) + C_4(\mu) + 3 C_5(\mu) +
C_6(\mu)~,\nnb \\
h \ga z,\hat s \dr \es - \frac{8}{9} \ln z +
\frac{8}{27} + \frac{4}{9} x -
\frac{2}{9} \ga 2 + x \dr \sqrt{\vel 1 - x \ver} \nnb \\
\cp \Bigg[ \Theta(1 - x)
\ga \ln \vel \frac{1  + \sqrt{1 - x}}{1  -  \sqrt{1 - x}} \ver - i \pi \dr
+ \Theta(x - 1) \, 2 \, \arctan \frac{1}{\sqrt{x - 1}} \Bigg], \nnb \\
h \ga 0,\hat s \dr \es - \frac{8}{27} - \frac{8}{9} \ln {m_b\over \mu}
 - \frac{4}{9} \ln \hat{s} + {4 \over 9} i \pi~,
\eea
and
\bea
\label{e10707}
Y_{LD}(\hat s) \es \frac{3}{\alpha^2} C^{(0)}(\mu)
\sum_{V_i = \psi \ga 1 s \dr, \cdots, \psi \ga 6 s \dr}
\ds{\frac{ \pi \ae_{i} \Gamma \ga V_i \rar \ell^+ \ell^- \dr M_{V_i} }
{\ga M_{V_i}^2 - \hat s m_b^2 - i M_{V_i} \Gamma_{V_i} \dr }}~,
\eea
The values of the phenomenological parameters $\ae_i$ are fixed from the 
analysis of $B \to K^* \ell^+ \ell^-$ decay and they are taken to be $\ae =
1.65$ and $\ae=2.36$ for the resonances $J/\psi$ and $\psi^\prime$,
respectively.

The charm loop brings further corrections to the radiative $b \to s \gamma$
transition, which modifies the Wilson coefficient $C_7^{eff}$. The Wilson
coefficient $C_7^{eff}$ can be written as \cite{R10725},
\bea
\label{nolabel}   
C_7^{eff} \es C_7(\mu) + C_7^\prime(\mu)~, \nnb
\eea
where
\bea
\label{e10708}
C_7^\prime(\mu) \es i \alpha_s \Bigg\{ {2\over 9} \eta^{14/23} \Bigg[F(x_t)
- 0.1687 \Bigg] - 0.03 C_2(\mu) \Bigg\}~, \nnb \\
F(x_t) \es { x_t (x_t^2 - 5 x_t -2) \over 8 (x_t-1)^2} +
{3 (x_t \ln x_t)^2 \over 4(x_t-1)^4}~, 
\eea
with $x_t=m_t^2/m_W^2$ and $\eta= \alpha_s(m_W)/\alpha_s(\mu)$.

As has already been noted, our main aim in this work is to analyze the $B
\to K_2(1430) \ell^+ \ell^-$ decay in a model independent way.The most
general, model independent local four--Fermi interaction is given in
\cite{R10715}, which might contribute to the considered decay. Explicit form
of the local four--Fermi interaction which describes $b \to s \ell^+ \ell^-$
transition can be written as,
\bea
\label{e10709}
{\cal M}_{new} \es {G_F \alpha \over \sqrt{2} \pi} V_{tb}V_{ts}^* \Bigg\{
C_{BR} \bar s_L i \sigma_{\mu\nu} \frac{q^\nu}{q^2} b_R \bar{\ell}
\gamma^\mu \ell + C_{SL} \bar s_R i \sigma_{\mu\nu} \frac{q^\nu}{q^2} b_L
\bar \ell \gamma^\mu \ell
+C_{LL} \bar{s}_L \gamma_\mu b_L \, \bar{\ell} \gamma_\mu \ell \nnb \\
\ar C_{LR} \bar s_L \gamma_\mu b_L \bar \ell_R \gamma^\mu \ell_R
+ C_{RL} \bar s_R \gamma_\mu b_R \bar \ell_L \gamma^\mu \ell_L + 
C_{RR} \bar s_R \gamma_\mu b_R \bar \ell_R \gamma^\mu \ell_R \nnb \\
\ar C_{LRRL} \bar s_L b_R \bar \ell_R \ell_L +
C_{LRLR} \bar s_L b_R \bar \ell_L \ell_R
+ C_{RLRL} \bar s_R b_L \bar \ell_R \ell_L \nnb \\
\ar C_{RLLR} \bar s_R b_L \bar \ell_L \ell_R +
C_T \bar s \sigma_{\mu\nu} b \bar \ell \sigma^{\mu\nu} \ell +
i C_{TE} \epsilon_{\mu\nu\alpha\beta} \bar s \sigma^{\mu\nu} b
\bar \ell \sigma^{\alpha\beta} \ell \Bigg\}~,
\eea
For the sake of simplicity we neglect the contribution of the tensor
interaction in further discussion.

Using Eqs. (\ref{e10703}) and (\ref{e10709}), the matrix element of the $b
\to s \ell^+ \ell^-$ transition, including the SM and new physics
contributions, can be written as,
\bea
\label{e10710}  
{\cal M} = {\cal M}_{SM} + {\cal M}_{new}~.
\eea
The matrix element for the exclusive $B \to K_2(1430) \ell^+ \ell^-$ decay
can be obtained from Eq. (\ref{e10710}) by sandwiching it between initial
and final states, i.e., 
\bea
\label{e10711}
\lla K_2(1430) (p,\varepsilon) \vel {\cal M} \ver B(p_B) \rra~.
\eea
Before giving definition of the matrix of quark operators between initial
and final meson states,  few words about the helicity states of the tensor
$K_2(1430)$ meson are in order. The polarizations
$\varepsilon_\lambda^{\mu\nu}$ with helicity $\lambda$ of the tensor meson
with mass $m$ and four--momentum $(E,0,0,p_z)$ moving along the z--axis can
be written in terms of the polarization vectors \cite{R10726}
\bea
\label{e10712}
\varepsilon_{(0)}^{*\mu} \es {1\over m} (p,0,0,E)~, \nnb \\
\varepsilon_{\pm}^{*\mu} \es {1\over \sqrt{2}} (0,\mp1,i,0)~,
\eea
in the following way,
\bea
\label{e10713}
\varepsilon_{\pm 2}^{*\alpha\beta} \es \varepsilon_{\pm}^\alpha
\varepsilon_{\pm}^\beta~,\nnb \\
\varepsilon_{\pm 1}^{*\alpha\beta} \es {1\over \sqrt{2}}
\Big[\varepsilon_{\pm}^\alpha \varepsilon_{0}^\beta +
\varepsilon_{0}^\alpha \varepsilon_{\pm}^\beta \Big]~, \nnb \\
\varepsilon_{0}^{*\alpha\beta} \es {1\over \sqrt{6}}
\Big[\varepsilon_{+}^\alpha \varepsilon_{-}^\beta + 
\varepsilon_{-}^\alpha \varepsilon_{+}^\beta \Big] +
\sqrt{2\over 3} \varepsilon_{0}^\alpha \varepsilon_{0}^\beta~.
\eea
It follows from the expression of the amplitude of the $b \to s \ell^+
\ell^-$ transition that, in order to obtain the matrix element for the
semileptonic $B \to K_2 (1430) \ell^+ \ell^-$ decay, the following matrix
elements are needed to be known,
\bea
\label{e10714}
&& \lla K_2(1430) (p,\varepsilon) \vel \bar{s} \gamma_\mu (1\pm \gamma_5) b \ver B(p_B) \rra~, \\ 
\label{e10715} 
&& \lla K_2(1430) (p,\varepsilon) \vel \bar{s} \sigma_{\mu\nu} q^\nu (1 + \gamma_5) b \ver B(p_B) \rra~, \\
\label{e10716} 
&& \lla K_2(1430) (p,\varepsilon) \vel \bar{s} (1\pm \gamma_5) b \ver B(p_B) \rra~.
\eea
The matrix element given in Eq. (\ref{e10714}) can be parametrized in terms
of the form factors as follows,
\bea
\label{e10717} 
&&\lla K_2 (p,\varepsilon) \vel \bar{s} \gamma_\mu (1\pm \gamma_5) b \ver
B(p_B) \rra = - \epsilon_{\mu\nu\alpha\beta} \varepsilon^{*\nu} p^\alpha
q^\beta {2 V(q^2) \over m_B+m_{K_2}}
\pm i \varepsilon_\mu^* (m_B+m_{K_2}) A_1(q^2) \nnb \\
&& \mp i(p_B+p)_\mu (\varepsilon^* q) 
{A_2(q^2) \over m_B+m_{K_2}} \mp i q_\mu {2 m \over q^2}  (\varepsilon^* q)
\left[A_3(q^2)-A_0(q^2)\right]~,\\ \nnb \\
\label{e10718}
&&\lla K_2 (p,\varepsilon) \vel \bar{s} \sigma_{\mu\nu} q^\nu (1 + \gamma_5) b \ver
B(p_B) \rra = 2 \epsilon_{\mu\nu\alpha\beta} \varepsilon^{*\nu} p^\alpha  
q^\beta T_1(q^2)
+ i \Big[\varepsilon_\mu^* (m_B^2-m_{K_2}^2) \nnb \\
&& - (p_B+p)_\mu (\varepsilon^* q)
\Big] T_2(q^2) +
i (\varepsilon^* q) \left[q_\mu - (p_B+p)_\mu {q^2 \over
m_B^2-m_{K_2}^2} \right]T_3(q^2)~,
\eea
where,
\bea
\label{nolabel}
\varepsilon_\lambda^{*\mu} \equiv  \varepsilon_\lambda^{*\mu\nu}
{q_\nu \over m_B}~.\nnb
\eea
The matrix element $\lla K_2 (p,\varepsilon) \vel \bar{s} (1\pm \gamma_5)
b \ver B(p_B) \rra$ can be obtained from Eq. (\ref{e10717}) by multiplying
it with $q_\mu$ and then using equation of motion. Neglecting strange quark
mass we get,
\bea
\label{e10719}
&&\lla K_2 (p,\varepsilon) \vel \bar{s} (1\pm \gamma_5) b \ver      
B(p_B) \rra = {1\over m_B}\Big\{ i (\varepsilon^* q) (m_B+m_{K_2}) A_1(q^2) \nnb \\
&& \pm (m_B-m_{K_2}) (\varepsilon^* q) A_2(q^2) \pm 2 m (\varepsilon^* q)
\left[ A_3(q^2) - A_0(q^2) \right] \Big\}~.
\eea
The following relation holds among the form factors $A_1(q^2)$, $A_2(q^2)$ and 
$A_3(q^2)$,
\bea
\label{e10720}
2 m A_3(q^2) = (m_B+m_{K_2}) A_1(q^2) - (m_B-m_{K_2}) A_2(q^2)~.
\eea
Using Eqs.(\ref{e10719}) and (\ref{e10720}), we get
\bea
\label{e10721}
\lla K_2 (p,\varepsilon) \vel \bar{s} (1\pm \gamma_5) b \ver B(p_B) \rra =
{1\over m_B} \left[ \pm 2 m (\varepsilon^* q) A_0(q^2) \right]~.
\eea

Using these definitions of the form factors, we get the following expression
for the decay amplitude of the $B \to K_2(1430) \ell^+ \ell^-$ channel,
\bea
\label{e10722}
{\cal M} \es {G_F \alpha \over 4 \sqrt{2} \pi} V_{tb} V_{ts}^* \Bigg\{
\bar{\ell} \gamma_\mu (1-\gamma_5) \ell \Big[ - 2 A_1
\epsilon_{\mu\nu\lambda\sigma} \varepsilon^{\nu *} p^\lambda q^\sigma - i
B_1 \varepsilon_\mu^* + i B_2 (\varepsilon^* q) (p_B+p)_\mu \nnb \\
\ar i B_3 (\varepsilon^* q) q_\mu \Big] + \bar{\ell} \gamma_\mu (1+\gamma_5)
\ell \Big[ - 2 C_1 \epsilon_{\mu\nu\lambda\sigma} \varepsilon^{\nu *}
p^\lambda q^\sigma - i D_1 \varepsilon_\mu^* + i D_2 (\varepsilon^* q)
(p_B+p)_\mu \nnb \\
\ar i D_3 (\varepsilon^* q) q_\mu \Big] + \bar{\ell} (1-\gamma_5) 
\ell \Big[i B_4 (\varepsilon^* q) \Big] + \bar{\ell} (1+\gamma_5)          
\ell \Big[i B_5 (\varepsilon^* q) \Big] \Bigg\}~.
\eea
Here
\bea          
\label{e10723}
A_1 \es (C_{LL}^{tot} + C_{RL}) \frac{V}{m_B+m_{K_2}} -
(C_{BR}+C_{SL}) \frac{T_1}{q^2} ~, \nnb \\
B_1 \es (C_{LL}^{tot} - C_{RL}) (m_B+m_{K_2}) A_1 -
(C_{BR}-C_{SL}) (m_B^2-m_{K_2}^2)
\frac{T_2}{q^2} ~, \nnb \\
B_2 \es \frac{C_{LL}^{tot} - C_{RL}}{m_B+m_{K_2}} A_2 -
(C_{BR}-C_{SL})
\frac{1}{q^2}  \left[ T_2 + \frac{q^2}{m_B^2-m_{K_2}^2} T_3 \right]~,
\nnb \\
B_3 \es 2 (C_{LL}^{tot} - C_{RL}) m \frac{A_3-A_0}{q^2}+
(C_{BR}-C_{SL}) \frac{T_3}{q^2} ~, \nnb \\
C_1 \es A_1 ( C_{LL}^{tot} \rar C_{LR}^{tot}~,~~C_{RL} \rar
C_{RR})~,\nnb \\
D_1 \es B_1 ( C_{LL}^{tot} \rar C_{LR}^{tot}~,~~C_{RL} \rar
C_{RR})~,\nnb \\
D_2 \es B_2 ( C_{LL}^{tot} \rar C_{LR}^{tot}~,~~C_{RL} \rar
C_{RR})~,\nnb \\
D_3 \es B_3 ( C_{LL}^{tot} \rar C_{LR}^{tot}~,~~C_{RL} \rar
C_{RR})~,\nnb \\
B_4 \es - 2 ( C_{LRRL} - C_{RLRL}) \frac{ m}{m_b} A_0 ~,\nnb \\
B_5 \es - 2 ( C_{LRLR} - C_{RLLR}) \frac{m}{m_b} A_0 ~,
\eea
where
\bea
\label{e10724}
C_{LL}^{tot} \es C_9^{eff} - C_{10} + C_{LL}~, \nnb \\
C_{LR}^{tot} \es C_9^{eff} + C_{10} + C_{LR}~, \nnb \\
C_{BR}       \es - 2 m_b C_7^{eff} +C_{BR}^\prime~.
\eea
It should be stressed at this point that, the matrix element of $B
\to K_2(1430) \ell^+ \ell^-$ decay is formally the same with that of the $B
\to V \ell^+ \ell^-$ decay ($V$ is $\rho$ or $K$ meson). But it is necessary
to keep in mind that form factors in both cases are different, and also, the
polarization vector $\varepsilon_\alpha^*$ which has the form,
\bea
\label{nolabel}
\varepsilon_\alpha^* = {\varepsilon_{\alpha\beta}^* q^\beta \over m_B}~, \nnb
\eea
is different from the polarization vector of the vector mesons.
 
Having obtained the matrix element $B \to K_2(1430) \ell^+ \ell^-$
decay, the next step in our analysis is to calculate the helicity amplitudes
for this decay. Using the helicity amplitude formalism presented in
\cite{R10727}, we get the following helicity amplitudes for the $B \to
K_2(1430) \ell^+ \ell^-$ decay,
\bea
\label{e10725}
M_{\pm}^{++} \es
\mp i {m_\ell \over m_B m_{K_2}} \vel \vec{p}_{K_2} \ver \sqrt{q^2} \sin\theta \Big[
\left( B_1 + D_1 \right)     
\mp 2 \vel \vec{p}_{K_2} \ver \sqrt{q^2} \left( A_1 + C_1 \right)
\Big]~, \nnb \\ \nnb \\
M_{\pm}^{+-} \es
i {( \mp 1 + \cos\theta) \over 2 m_B m_{K_2} } \vel \vec{p}_{K_2} \ver q^2
\Big\{ \mp (1-v) B_1 \mp (1+v) D_1 \nnb \\
\ar 2 \vel \vec{p}_{K_2} \ver \sqrt{q^2} \Big[ (1-v) A_1 + (1+v) C_1
\Big] \Big\}~, \nnb \\ \nnb \\
M_{\pm}^{-+} \es 
i {( \pm 1 + \cos\theta) \over 2 m_B m_{K_2} } \vel \vec{p}_{K_2} \ver q^2
\Big\{ \mp (1+v) B_1 \mp (1-v) D_1 \nnb \\
\ar 2 \vel \vec{p}_{K_2} \ver \sqrt{q^2} \Big[ (1+v) A_1 + (1-v) C_1
\Big] \Big\}~, \nnb \\ \nnb \\
M_{\pm}^{--} \es 
\pm i {m_\ell \over m_B m_{K_2}} \vel \vec{p}_{K_2} \ver \sqrt{q^2} \sin\theta \Big[
\left( B_1 + D_1 \right)
\mp 2 \vel \vec{p}_{K_2} \ver \sqrt{q^2} \left( A_1 + C_1 \right)
\Big]~, \nnb \\ \nnb \\
M_{0}^{++}   \es i {\sqrt{2/3} \over m_B m_{K_2}^2} \vel \vec{p}_{K_2} \ver \sqrt{q^2}\Big\{
2 m_\ell \Big[ E_{K_2} \cos\theta \left(B_1 + D_1\right) +
\vel \vec{p}_{K_2} \ver \left( B_1 - D_1\right) \Big] \nnb \\
\ek 2 m_\ell \vel \vec{p}_{K_2} \ver \sqrt{q^2} \Big[ \left( B_2 - D_2\right)
\left( E_B + E_{K_2}\right) +
2 \vel \vec{p}_{K_2} \ver \cos\theta \left( B_2 + D_2\right) \Big] \nnb \\
\ek \vel \vec{p}_{K_2} \ver q^2 \Big[ 2 m_\ell \left( B_3 - D_3\right) +                      
(1+v) B_4 - (1-v) B_5 \Big] \Big\}~, \nnb \\ \nnb \\
M_{0}^{+-}   \es 
i {\sqrt{2/3} \over m_B m_{K_2}^2 } \vel \vec{p}_{K_2} \ver q^2 \sin\theta 
\Big\{ - E_{K_2} \Big[ B_1 (1-v) + 
D_1 (1+v) \Big] \nnb \\
\ar 2 \vel \vec{p}_{K_2} \ver^2 \sqrt{q^2} \Big[ B_2 (1-v) + 
D_2 (1+v) \Big] \Big\}~, \nnb \\ \nnb \\
M_{0}^{-+}   \es
i {\sqrt{2/3} \over m_B m_{K_2}^2 } \vel \vec{p}_{K_2} \ver q^2 \sin\theta  
\Big\{ - E_{K_2} \Big[ B_1 (1+v) + 
D_1 (1-v) \Big] \nnb \\
\ar 2 \vel \vec{p}_{K_2} \ver^2 \sqrt{q^2} \Big[ B_2 (1+v) + 
D_2 (1-v) \Big] \Big\}~, \nnb \\ \nnb \\
M_{0}^{--}   \es i {\sqrt{2/3} \over m_B m_{K_2}^2} \vel \vec{p}_{K_2} \ver
\sqrt{q^2}\Big\{
- 2 m_\ell \Big[ E_{K_2} \cos\theta \left( B_1 + D_1\right) +
\vel \vec{p}_{K_2} \ver \left( B_1 - D_1\right) \Big] \nnb \\
\ek 2 m_\ell \vel \vec{p}_{K_2} \ver \sqrt{q^2} \Big[ \left( B_2 - D_2\right)
\left( E_B + E_{K_2}\right) -
2 \vel \vec{p}_{K_2} \ver \cos\theta \left( B_2 + D_2\right) \Big] \nnb \\
\ek \vel \vec{p}_{K_2} \ver q^2 \Big[ 2 m_\ell \left( B_3 - D_3 \right) +
(1-v) B_4 - (1+v) B_5 \Big] \Big\}~.
\eea
Here superscripts and subscripts denote the helicities of the leptons and
$K_2$ meson, respectively. In Eq. (\ref{e10726}) we have,
\bea
\label{nolabel}
\lambda(m_B^2,q^2,m_{K_2}^2) \es m_B^4 + q^4 + m_{K_2}^4 - 
2 m_B^2 q^2 - 2 m_B^2 m_{K_2}^2 - 2 q^2 m_{K_2}^2~, \nnb \\
q^2 \es (p_B - p_{K_2})^2 = (p_1+p_2)^2~, \nnb \\
v \es \sqrt{1-{4 m_\ell^2\over q^2}}~, \nnb \\
\vel \vec{p}_{K_2}\ver \es {\lambda^{1/2}(m_B^2,q^2,m_{K_2}^2) 
\over 2 m_B}~, \nnb
\eea
and $m_\ell$ is the lepton mass, $\theta$ is the angle between $K_2$ and
$\ell^-$ lepton.

It should be noted here that the $\pm 2$ helicity states of the tensor meson
give no contribution to the helicity amplitudes. This is due to the fact
that in the CM of leptons only time component of $q^2$ is different from
zero, and therefore $\varepsilon_\pm^{*\alpha} q_\alpha = 0$.

As a result of some calculation, we obtain the differential decay width in
terms of the helicity amplitudes as follows,
\bea
\label{e10726}
{d \Gamma \over dq^2 \, d\!\cos\theta} \es {G_F^2 \alpha^2 \vel V_{tb} V_{ts}^*
\ver^2 \over 2^{14} \pi^5 m_B^3} \lambda^{1/2} v \Bigg\{
\vel M_+^{+-} \ver^2 + \vel M_-^{+-} \ver^2 + \vel M_+^{++} \ver^2
+ \vel M_-^{++} \ver^2 + \vel M_+^{-+} \ver^2 \nnb \\
\ar \vel M_-^{-+} \ver^2 + \vel M_+^{--} \ver^2
+ \vel M_-^{--} \ver^2 + \vel M_0^{++} \ver^2 + \vel M_0^{+-} \ver^2
+ \vel M_0^{-+} \ver^2 + \vel M_0^{--} \ver^2 \Bigg\}~.
\eea 

We can now proceed to calculate the quantities 
${\ds {\Gamma_+ \over \Gamma_-}}$ and
${\ds {\Gamma_L \over \Gamma_T} = {\Gamma_0 \over \Gamma_+ + \Gamma_-}}$ 
(here, subscripts $+,-,0$ correspond to
the tensor meson helicities), lepton forward--backward asymmetry and
longitudinal polarization of the final lepton. These quantities can all be
measured in experiments. Since these quantities all involve ``new" Wilson
coefficients, they might be very sensitive to new physics.

The expressions for the quantities $\Gamma_\pm$ and $\Gamma_0$ can easily be
obtained from Eq. (\ref{e10726}), which can be written as,
\bea          
\label{e10727}
\Gamma_\pm \es {G_F^2 \alpha^2 \vel V_{tb} V_{ts}^*  
\ver^2 \over 2^{14} \pi^5 m_B^3} \int_{4 m_\ell^2}^{(m_B-m_{K_2})^2} 
dq^2 v \lambda^{1/2} \int_{-1}^{+1} d\!\cos\theta \Bigg\{\vel M_\pm^{++} \ver^2
+ \vel M_\pm^{--} \ver^2 \nnb \\
\ar \vel M_\pm^{-+} \ver^2 + \vel M_\pm^{+-} \ver^2
\Bigg\}~, \\
\label{e10728}
\Gamma_0 \es {G_F^2 \alpha^2 \vel V_{tb} V_{ts}^*
\ver^2 \over 2^{14} \pi^5 m_B^3} \int_{4 m_\ell^2}^{(m_B-m_{K_2})^2}
dq^2 v \lambda^{1/2}\int_{-1}^{+1} d\!\cos\theta \Bigg\{\vel M_0^{++} \ver^2
+ \vel M_0^{--} \ver^2 \nnb \\
\ar \vel M_0^{-+} \ver^2 + \vel M_0^{+-} \ver^2  
\Bigg\}~.
\eea

The differential forward--backward asymmetry of the state final lepton can be
obtained from Eq. (\ref{e10726}) in the following way,
\bea          
\label{e10729} 
{d{\cal A}_{FB} \over d q^2} = \int_{0}^{+1} d\!\cos\theta {d\Gamma \over
dq^2 \, d\!\cos\theta } - \int_{-1}^{0} d\!\cos\theta {d\Gamma \over   
dq^2 \, d\!\cos\theta }~.
\eea  

At the end of this section we present the expression for the longitudinal
polarization of the final state lepton, which also might be very useful for
establishing new physics beyond the SM. The expression for the longitudinal
polarization of the final state lepton can easily be calculated from Eq.
(\ref{e10726}), which has the following form (see also \cite{R10728}),
\bea          
\label{e10730}
{\cal P}_L = {
\ds \int_{4 m_\ell^2}^{(m_B-m_{K_2})^2} 
dq^2 v \lambda^{1/2} \int_{-1}^{+1} d\!\cos\theta \Big[ \chi_1 -
\chi_2\Big]
\over
\ds \int_{4 m_\ell^2}^{(m_B-m_{K_2})^2} 
dq^2 v \lambda^{1/2} \int_{-1}^{+1} d\!\cos\theta \Big[ \chi_1 +
\chi_2\Big]
             }~,
\eea
where
\bea
\label{nolabel}
\chi_1 \es \vel M_-^{-+} \ver^2 + \vel M_+^{-+} \ver^2 +
\vel M_-^{--} \ver^2 + \vel M_+^{--} \ver^2 +
\vel M_0^{-+} \ver^2 + \vel M_0^{--} \ver^2~, \nnb \\
\chi_2 \es \vel M_-^{+-} \ver^2 + \vel M_+^{+-} \ver^2 + 
\vel M_-^{++} \ver^2 + \vel M_+^{++} \ver^2 +
\vel M_0^{+-} \ver^2 + \vel M_0^{++} \ver^2~. \nnb
\eea

\section{Numerical analysis}

In this section we investigate the dependence of the physical quantities
mentioned in section 2, on the new Wilson coefficients. The main input
parameters in our calculations are the new Wilson coefficients and the form
factors responsible for the $B \to K_2$ transition. We use the results of
\cite{R10729} for the form factors, which are calculated within the QCD sum
rules method. The $q^2$ dependence of all form factors are described by the
following formula,

\bea
\label{nolabel}
F_i(q^2) = \frac{F_i(0)}{\ds 1-a_i \left( {q^2 \over m_B^2}\right) + b_i
\left({q^2 \over m_B^2}\right)^2}~. \nnb
\eea
The values of parameters $F_i(0)$, $a_i$ and $b_i$ for different form factors
are in Table 1 (this table is taken from \cite{R10729}).

\begin{table}[h]
\renewcommand{\arraystretch}{1.5}
\addtolength{\arraycolsep}{3pt}
$$
\begin{array}{|l|ccc|}   
\hline
       &      F(0)     & a    & b    \\ \hline
A_0    & 0.25 \pm 0.04 & 1.57 & 0.10 \\
A_1    & 0.14 \pm 0.02 & 1.21 & 0.52 \\
A_2    & 0.05 \pm 0.02 & 1.32 & 14.9 \\
V      & 0.16 \pm 0.02 & 2.08 & 1.50 \\
T_1    & 0.14 \pm 0.02 & 2.08 & 1.50 \\
T_2    & 0.14 \pm 0.02 & 1.22 & 0.35 \\
T_3    & 0.01_{-0.01}^{+0.02} & 9.91 & 276  \\ \hline
\end{array}
$$       
\caption{$B$ meson decay form factors in a three--parameter fit.}
\renewcommand{\arraystretch}{1}
\addtolength{\arraycolsep}{-3pt}
\end{table}

The Wilson coefficients $C_7^{eff}$ and $C_9^{eff}$ which we use in our
analysis are given in Eqs. (\ref{e10704}) and (\ref{e10708}) with
$C_9=4.253$ and $C_7=-0.311$ at $\mu=m_b$ scale and $C_{10}=-4.546$.

As has already been noted, other input parameters needed are the new Wilson
coefficients. A systematic analysis of the $B \to K^* \ell^+ \ell^-$ decay
is carried out in \cite{R10720} using the most general, model independent 
Hamiltonian to find the constraints on the new Wilson coefficients. In
accordance with the result of \cite{R10720}, we will vary the vector type
new Wilson coefficients in between $-C_{10}$ and $+C_{10}$. For the scalar
type operators the following constraint is obtained in \cite{R10720},
\bea
\label{nolabel}
\vel C_{LRLR} \ver^2 + \vel C_{LRRL} \ver^2 &\le& 0.44~,~~~\mbox{(from $B \to
\mu^+ \mu^-$ )}~, \nnb \\
 {1\over 2} \left(\vel C_{LRLR} \ver^2 + \vel C_{LRRL} \ver^2\right) &\le& 
45~,~~~\mbox{(from $B \to X_s\mu^+ \mu^-$ )}~, \nnb
\eea

In our numerical calculations we shall use rather a broader range for the
scalar type Wilson coefficients, i.e., we assume that they also vary between
$-C_{10}$ and $+C_{10}$.

In Figs. (1) and (2) we present the dependence of $\Gamma_L/\Gamma_T$ on the
new Wilson coefficients for $\mu$ and $\tau$ lepton channels, respectively.

We see from Fig. (1) that $\Gamma_L/\Gamma_T$ is most sensitive to the
vector type coefficients, and practically insensitive to scalar type
interaction. The ratio $\Gamma_L/\Gamma_T$ becomes smaller (larger) in the
presence of the Wilson coefficients $C_{LL}$ and  $C_{LR}$ compared
to the SM value when these coefficients get negative (positive) values. This
situation is to the contrary when $C_{RL}$ takes role in the numerical
calculations.

For the $\tau$ channel, the ratio $\Gamma_L/\Gamma_T$ is strongly dependent
on the coefficients $C_{LR}$ and $C_{RR}$, and considerable change happens
also in the presence of the scalar interaction. We observe from the
relevant figure that the ratio changes several times compared to the SM case.

We further study also the dependence of the asymmetry parameter $\alpha = \ds{
\Gamma_- - \Gamma_+ \over \Gamma_- + \Gamma_+}$ on the new Wilson coefficients.
The result of this analysis can be summarized briefly as follows: This ratio
is sensitive only to the vector type new Wilson coefficients, but insensitive
to the presence of the scalar interactions. The parameter $\alpha$ exhibits
strong dependence on the coefficients $C_{RL}$ and $C_{RR}$.

We present in Fig. (3) the dependence of the forward--backward asymmetry for
the $\mu$ channel on $q^2$, when $C_{LL}$ is taken into account. We deduce
from this figure that, there is quite a significant shift in the zero position
of zero of ${\cal A}_{FB}$. The zero position is shifted to the right (left)
for the negative (positive) values of the coefficient $C_{LL}$ compared to
the SM value.

In further numerical analysis we also investigate the dependence the
forward--backward asymmetry ${\cal A}_{FB}$ on other new Wilson coefficients
for the $\mu$ and $\tau$ channels, and the outcome of these results can be 
summarized as follows:\\

\underline{In the case of $\mu$--channel}

\begin{itemize}

\item The zero position of ${\cal A}_{FB}$ is shifted to the right (left)
compared to the SM prediction when $C_{LR}$ gets positive (negative) values,
similar to the case when $C_{LL}$ is present.

\item The zero position of ${\cal A}_{FB}$ is insensitive to the presence of
the coefficients $C_{RL}$ and $C_{RR}$, and also to all scalar interaction
coefficients.

\item the maximum value of ${\cal A}_{FB}$ is realized in the presence of
the coefficient $C_{RL}$. 

\end{itemize}

\underline{In the case of $\tau$--channel}

\begin{itemize}

\item The zero position of ${\cal A}_{FB}$ is shifted to the right (left)   
when the Wilson coefficients $C_{LL}$ and $C_{LR}$ runs over negative (positive)
values. The situation here is different compared to the $\mu$ case.

\item The zero position of ${\cal A}_{FB}$ is insensitive to the presence of
all remaining coefficients.

\item The behavior of ${\cal A}_{FB}$ is very sensitive to the coefficients
$C_{RL}$, $C_{LRRL}$ and $C_{RLRL}$ to the variation in $q^2$ in the range
$q^2 > 13.8~GeV^2$. It is observed that the value of ${\cal A}_{FB}$ is
magnified 2--4 times compared to that of the SM case. Therefore the measurement
of the ${\cal A}_{FB}$ can be very useful for establishing new physics
beyond the SM.

\end{itemize}

In Figs. (4) and (5) we present the dependence of the final $\mu$ and $\tau$
leptons longitudinal polarizations on the new Wilson coefficients. We
observe from these figures that,\\

\underline{In the case of $\mu$--channel}

\begin{itemize}

\item
$P_L$ is sensitive to the existence of all new Wilson coefficients, except
the coefficients $C_{RLLR}$ and $C_{LRLR}$. The dependence of $P_L$ on 
$C_{LL}\,(C_{RL})$ has the tendency to increase (decrease) in the region 
$-C_{10} \le  C_{RL}\, (C_{LL}) \le
C_{10}$. For all other coefficients $P_L$ increase firstly in the region 
from $-C_{10}$ to zero (this region is from $-C_{10}$ to two for the coefficient
$C_{LR}$) and then decreases when the new Wilson coefficients vary in the
region from zero to $C_{10}$.

\end{itemize}

\underline{In the case of $\tau$--channel}

\begin{itemize}

\item In this channel $P_L$ exhibits strong dependence on the coefficients
$C_{LR}$, $C_{RR}$ and also on the scalar interaction coefficients
$C_{LRRL}$ and $C_{RLRL}$. Therefore the measurement of the
longitudinal polarization of the leptons can be quite informative about the
nature and the confirmation of the new physics beyond the SM.

\end{itemize}

\section{Conclusion}

In this work the sensitivity of the physically measurable quantities, such
as $\Gamma_L/\Gamma_T$, ${\cal A}_{FB}$ and the final lepton polarization
for the $B \to K_2 \ell^+ \ell^-$ decay is investigated using the most
general, model independent four--fermion interaction. It is observed that
the ratio $\Gamma_L/\Gamma_T$ is quite sensitive to the new Wilson
coefficients $C_{LL}$, $C_{LR}$ and $C_{RR}$ for the $B \to K_2 \mu^+
\mu^-$ channel, while for the $B \to K_2 \tau^+ \tau^-$ channel this ratio
is strongly dependent on the coefficients $C_{LR}$ and $C_{RR}$. This ratio
is rather weakly dependent on the scalar interaction coefficients.

We also studied in detail the dependence of the forward--backward asymmetry for
both channels on $q^2$. It is found that the zero position of ${\cal
A}_{FB}$ is shifted to right or left compared to its SM value. We also show
that the value of ${\cal A}_{FB}$ for the $\tau$ channel is quite sensitive
to the existence of scalar type interactions. The longitudinal polarization
of the leptons shows sensitivity to all new Wilson coefficients, except the
coefficient $C_{LRLR}$.

Measurement of these quantities can give invaluable information, not only
about the existence of new physics, but also about the signs of the new
Wilson coefficients.

\newpage

\section*{Figure captions}
{\bf Fig. (1)} The dependence of the ratio of the decay widths when $K_2$
meson is longitudinally and transversally polarized on the new Wilson
coefficients for the $B \to K_2 \mu^+ \mu^-$ decay. \\ \\
{\bf Fig. (2)} The same as in Fig. (1), but for the $B \to K_2 \tau^+ \tau^-$
decay. \\ \\
{\bf Fig. (3)} The dependence of the forward--backward asymmetry on $q^2$ at
several fixed values of the Wilson coefficient $C_{LL}$ for the $B \to K_2
\mu^+ \mu^-$ decay. \\ \\
{\bf Fig. (4)} The dependence of the longitudinal lepton polarization of
the $\mu$--lepton on the new Wilson coefficients. \\ \\
{\bf Fig. (5)} The same as in Fig. (4), but for the
$\tau$--lepton.

\newpage

\begin{figure}
\vskip 2.5 cm
    \includegraphics{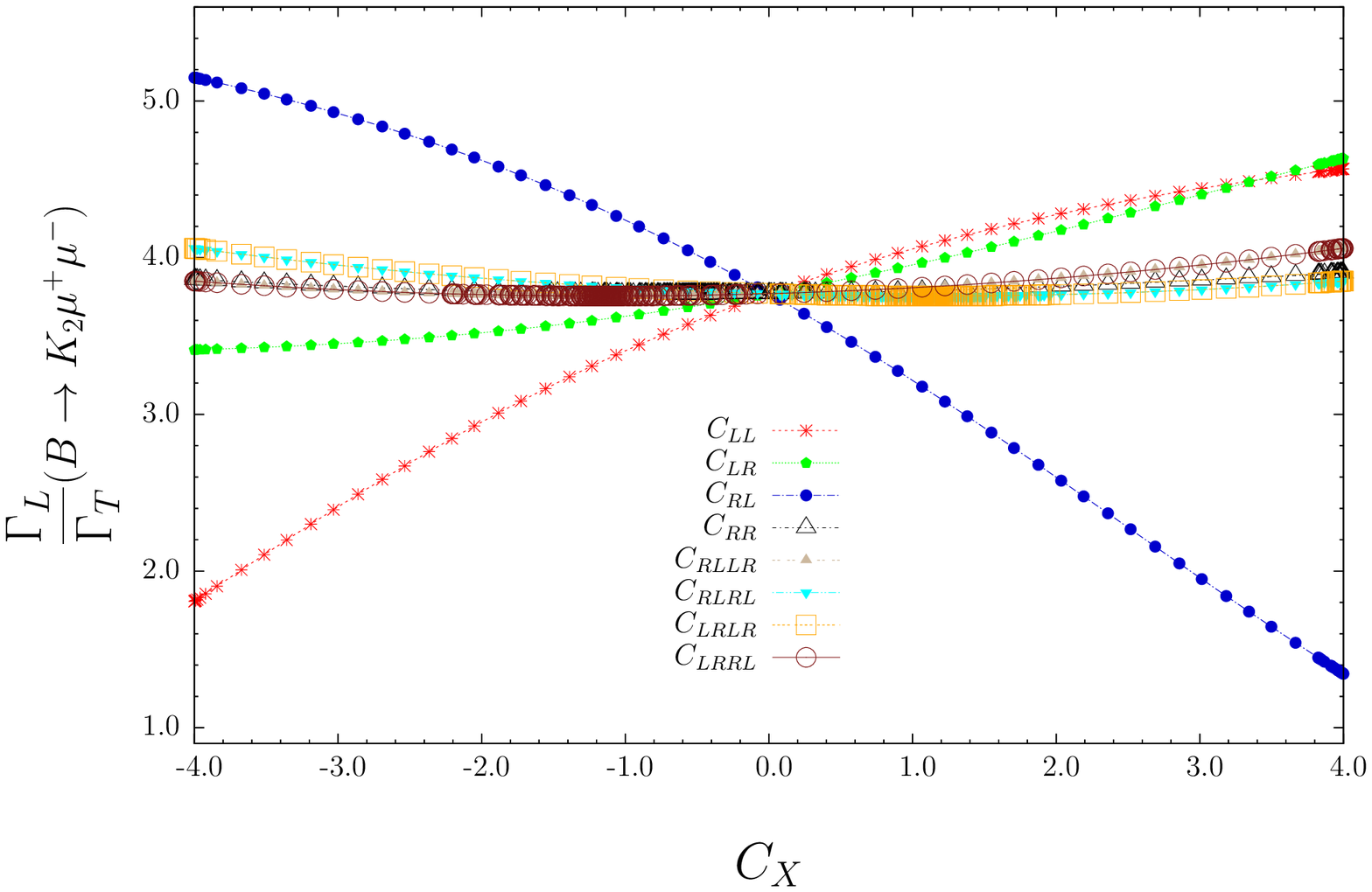}
\vskip 7.8cm
\caption{}
\end{figure}

\begin{figure}
\vskip 2.5 cm
    \includegraphics{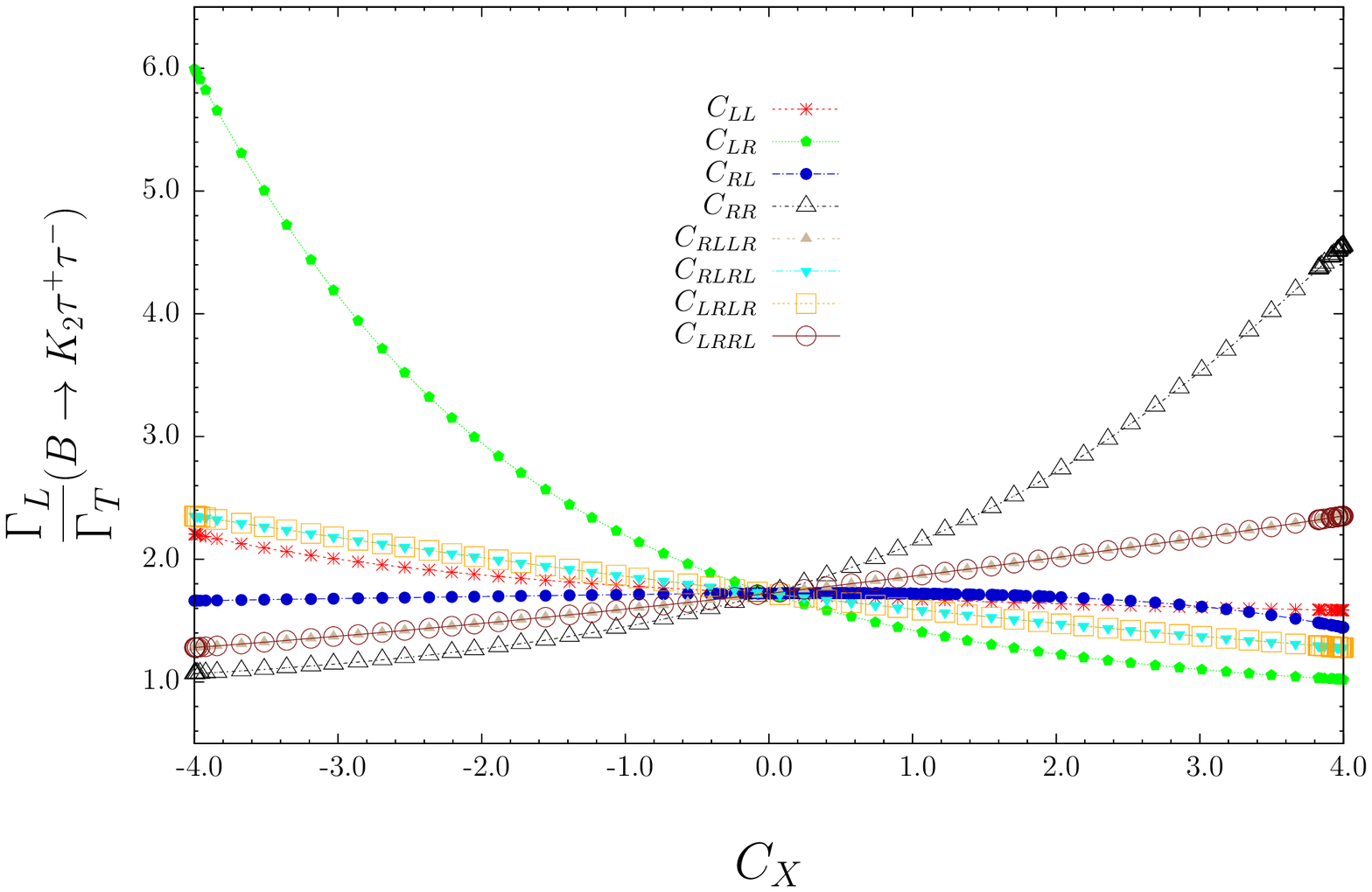}
\vskip 7.8 cm
\caption{}
\end{figure}

\begin{figure}
\vskip 2.5 cm
    \includegraphics{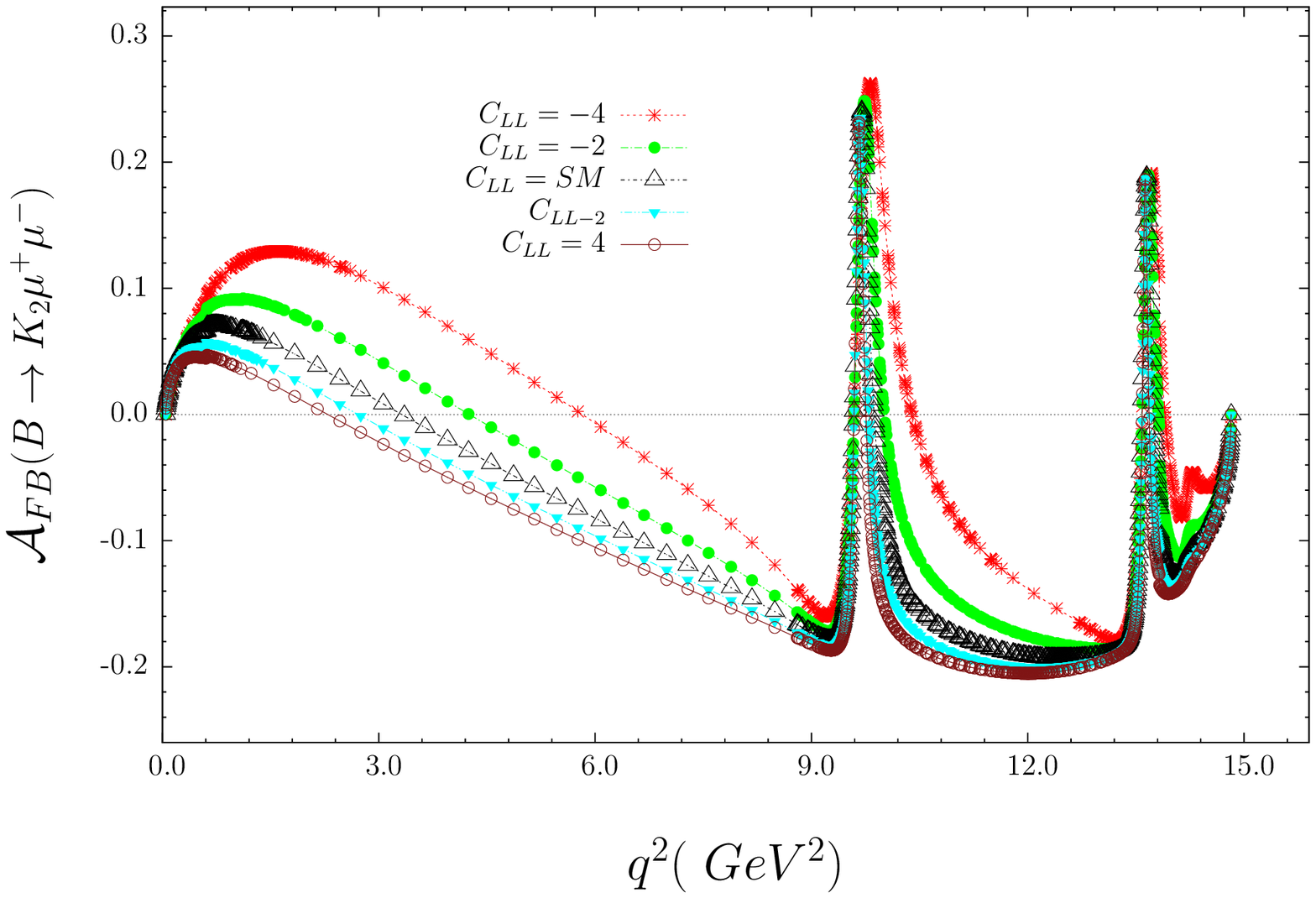}
\vskip 7.8cm
\caption{}
\end{figure}

\begin{figure}
\vskip 2.5 cm
    \includegraphics{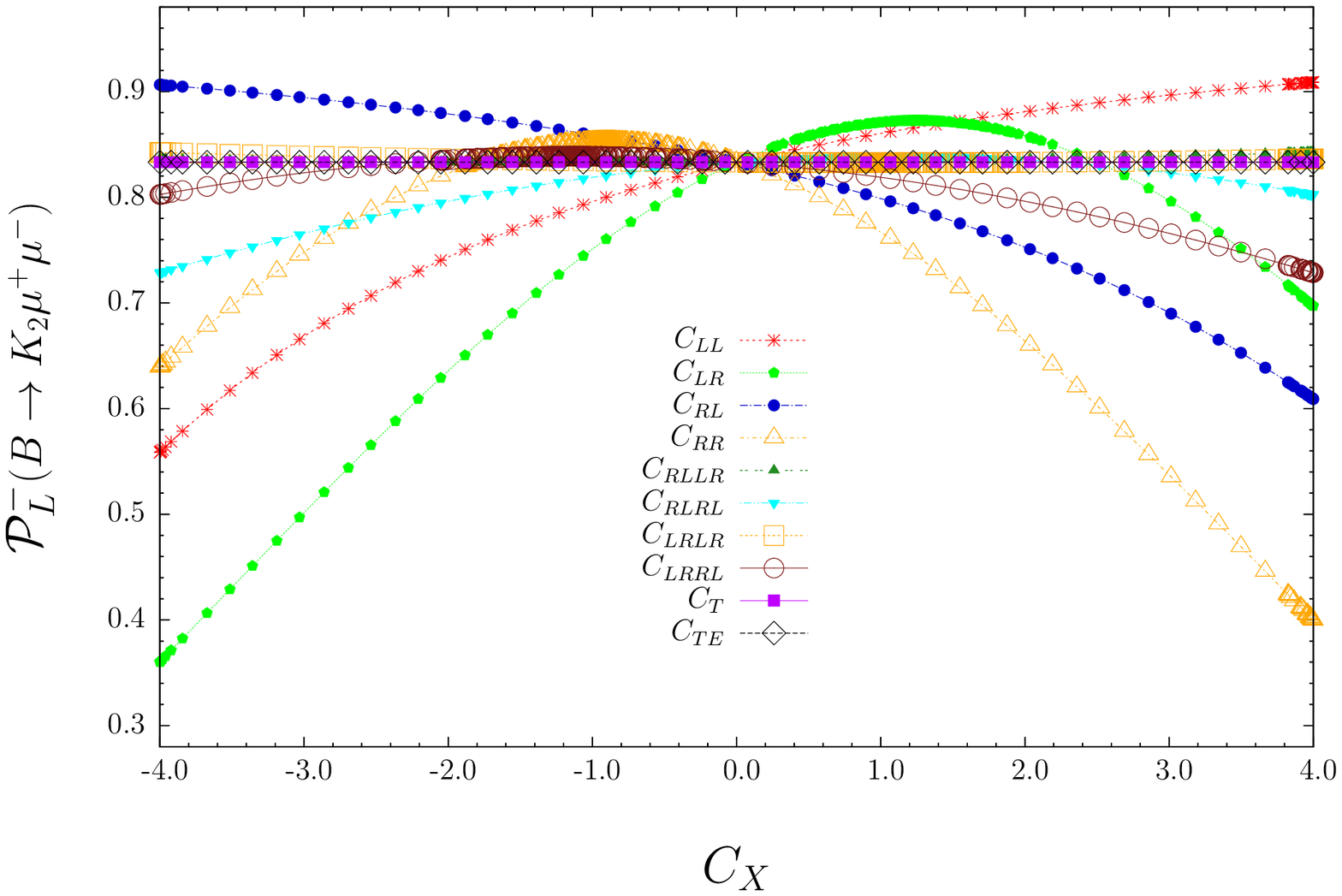}
\vskip 7.8 cm
\caption{}
\end{figure}

\begin{figure}
\vskip 2.5 cm
    \includegraphics{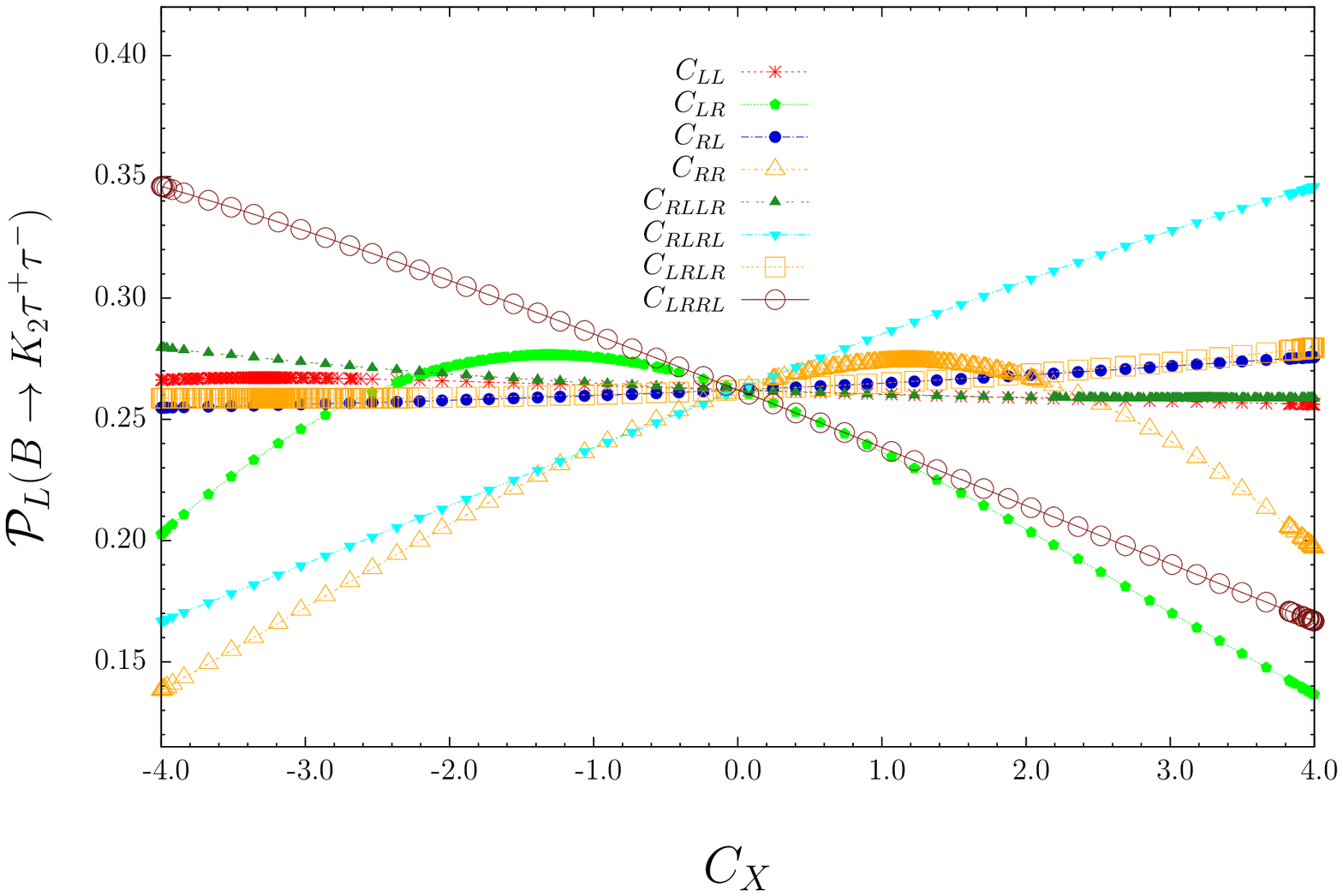}
\vskip 7.8 cm
\caption{}
\end{figure}

\end{document}